\documentclass{article}
\usepackage{graphicx}
\title{The IMF from Low to High Redshift}

\author{Laura Greggio and Alvio Renzini \thanks{e-mail:
    laura.greggio@oapd.inaf.it, alvio.renzini@oapd.inaf.it}\\
\small INAF-Osservatorio Astronomico di Padova\\
\small Vicolo dell'Osservatorio 5, I-35122, Padova, Italy}

\def\xi{\hbox{$X_{\rm i}$}}
\catcode`\@=11
\def\gsim{\ifmmode{\mathrel{\mathpalette\@versim>}}
   \else{$\mathrel{\mathpalette\@versim>}$}\fi}
\def\lsim{\ifmmode{\mathrel{\mathpalette\@versim<}}
    \else{$\mathrel{\mathpalette\@versim<}$}\fi}
\def\@versim#1#2{\lower 2.9truept \vbox{\baselineskip 0pt \lineskip 
    0.5truept \ialign{$\m@th#1\hfil##\hfil$\crcr#2\crcr\sim\crcr}}}
\catcode`\@=12

\def\lb{\hbox{$L_{\rm B}$}}

\def\lsun{\hbox{$L_\odot$}}
\def\zsun{\hbox{$Z_\odot$}}
\def\zf{\hbox{$z_{\rm F}$}}

\def\t9{\hbox{$t_9$}}

\def\zsun{\hbox{$Z_\odot$}}

\def\pn{\par\noindent}

\def\mfe{\hbox{$M_{\rm Fe}$}}
\def\mto{\hbox{$M_{\rm TO}$}}

\def\m*{\hbox{$M_*$}}

\def\mfe{\hbox{$M_{\rm Fe}$}}

\def\ho{\hbox{$H_\circ$}}
\def\h50{\hbox{$\ho /50$}}
\def\h70{\hbox{$h_{70$}}}
\def\y1{\hbox{${\rm yr}^{-1}$}}
\def\pn{\par\noindent}

\def\msun{\hbox{$M_\odot$}}
\date{}
\begin{document}
\renewcommand\thefigure{8.\arabic{figure}}
\renewcommand\theequation{8.\arabic{equation}}
 \maketitle
\begin{abstract}
From time to time, and quite more frequently in recent years, claims
  appear favoring a variable Initial Mass Function (IMF), one way or
  another, either in time or space.  In this chapter we add our two
  pennies of wisdom, illustrating how  the IMF affects various properties of
  galaxies and galaxy clusters. We start by showing that even
  relatively  small variations of the IMF slope have large effects on the
  demography of stellar populations, moving the bulk of the stellar
  mass from one end to
  the other of the distribution.
  We then point out how the slope of the IMF
  in different mass ranges controls specific major properties of
  galaxies and clusters.  The slope of the IMF below $\sim
  1\;\msun$ controls the $M/L$ ratio of local ellipticals, whereas the
  slope between $\sim 1$ and $\sim 1.4\;\msun$ controls the evolution
  with redshift of such ratio, hence of the fundamental plane of
  elliptical galaxies.  Similarly, the slope between $\sim 1$ and
  $\sim 40\;\msun$ drives the ratio of the global metal mass in  clusters of
  galaxies to their total luminosity. While we believe that it is perfectly legitimate to
  entertain the notion that the IMF may not be universal, our message
  is that when proposing IMF variations to ease a specific problem then
  one should not neglect to explore the full consequences of the
  invoked variations.

This paper is integrally reproduced  from Chapter 8 of the book by
L. Greggio \& A. Renzini: {\it Stellar Populations. A User Guide from
Low to High Redshift}  (2011, Wiley-VHC Verlag-GmbH \& Co., ISBN
9783527409181), whose index is also appended.
\end{abstract}
\newpage
\begin{Large}
$\,$
\vspace{2.2 truecm}

\pn
\textbf{8}
\vspace{0.3 truecm}
\pn
\textbf{The IMF from Low to High Redshift}
\end{Large}
\vspace{1 truecm}
\pn
At all redshifts much of galaxy properties depend on the IMF,
including mass-to-light ratios, derived galaxy masses and star
formation rates, the rate of the luminosity evolution of the
constituent stellar populations, the metal enrichment, and so on . With so
many important issues at stake, we still debate as to whether the IMF
is universal, that is, the same in all places and at all cosmic times, or
whether it depends on local conditions such as the intensity of star
formation (starburst vs. steady star formation), or on cosmic time,
for example, via the temperature of the microwave background.  As is well known,
we do not have anything close to a widely accepted theory of the IMF,
and this situation is likely to last much longer than
desirable. Again, star formation is an extremely complex
(magneto)hydrodynamical process, indeed much more complex than stellar
convection or red giant mass loss, for which we already noted the
absence of significant theoretical progress over the last 40-50
years. Thus, the IMF is parametrized for example., as one or more power laws
or as a lognormal distribution, and the parameters are fixed from
pertinent observational constraints.  Wherever the IMF has been
measured from statistically significant stellar counts, a {\it
Salpeter} IMF has been found, that is, $\phi(M)\propto M^{-s}$, with
$s=1+x\simeq 2.35$, however with a flattening to $s\simeq 1.3$ below
$\sim 0.5\,\msun$. Specifically, where possible, this has been proved
for stellar samples including the solar vicinity, open and globular
clusters in the Galaxy and in the Magellanic Clouds, actively
starbursting regions, as well as the old galactic bulge. Nevertheless,
this does not prove the universality of the IMF, as --with one
exception-- more extreme environments have not been tested yet in the
same direct fashion. The exception is represented by the very center
the of the Milky Way, in the vicinity of the supermassive black hole,
a very extreme environment indeed, where very massive stars seem to
dominate the mass distribution.  In this chapter we discuss a few
aspects of the IMF, using some of the stellar population tools that
have been illustrated in the previous chapters, and exploring how
specific integral properties of stellar populations depend on the IMF
slope in specific mass intervals. In particular, the dependence on the
IMF of the mass-to-light ratio of stellar populations is illustrated,
along with its evolution as stellar populations passively age. Then
the \textit{M/L} ratios of synthetic stellar populations, and their time
evolution, are compared to the dynamical \textit{M/L} ratios of local
elliptical galaxies, as well as to that of ellipticals up to redshift
$\sim 1$ and beyond. These comparisons allow us to set some constraint
on the low-mass portion of the IMF, from $\sim 0.1$ to $\sim
1.4\,\msun$. A strong constraint of the IMF slope from $\sim 1\,\msun$
up to $\sim 40\,\msun$ and above is then derived from considering the
metal content of clusters of galaxies together with their integrated
optical luminosity.

\vspace{0.65 truecm}
\pn
\textbf{8.1}
\pn
\textbf{How the IMF Affects Stellar Demography}
\vspace{0.5 truecm}
\pn
For a fixed amount of gas turned into stars, different IMFs obviously
imply different proportions of low mass and high mass stars. This is
illustrated in Figure~8.1 showing three different IMFs, all with
the same slope below $0.5\,\msun$, that is $s=1+x=1.35$, and three
different slopes above:

\begin{equation}
\begin{tabular}{ r l }
\(\phi(M) $  & \(= A\,M^{-s}\quad\quad\quad\quad\quad\quad\;\;\;  
              {\rm for\;\;} M\geq 0.5\,\msun\)
\\
 & \(= 0.5^{1.3-s}A\, M^{-1.3}\quad\quad\;\;\; {\rm for} \;\, \,M \leq 0.5\,\msun \)$$
 \end{tabular}
\label{eq_imf}
\end{equation}
where the factor $ 0.5^{1.3-s}$ ensures the continuity of
the IMF at $M=0.5\,\msun$. The normalization of the three IMFs corresponds to
a fixed amount ${\cal M}$ of gas turned into stars, that is, for fixed
\begin{equation}
{\cal M} = \int_{0.1}^{120}M\phi(M)dM.
\end{equation}
Here the case $s=2.35$ corresponds to the Salpeter-diet IMF already
encountered in previous chapters. Thick lines in Figure~8.1b  show the
\begin{figure}[h!b]
\includegraphics[width=12cm]{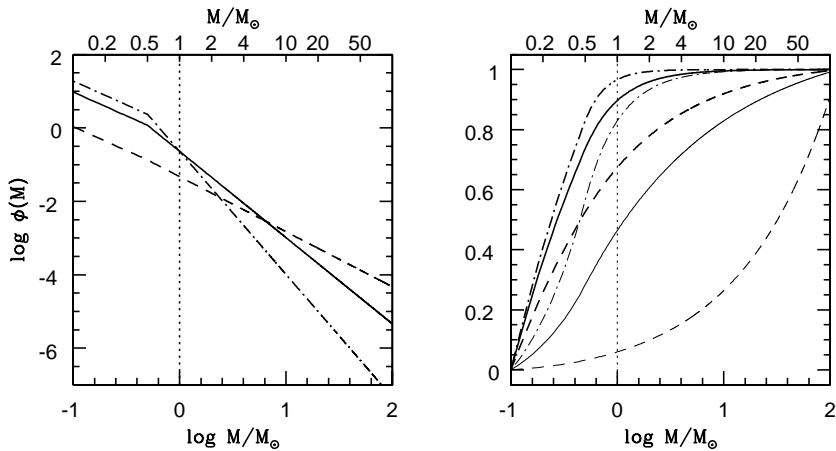}
\vskip -0.3 truecm
\caption{\footnotesize Left: three different IMFs normalized to have the same total
stellar mass of 1 \msun.
Below $0.5\,\msun$ all three IMFs have the same slope
$s=$1.3 and above it $s=$1.5, 2.35 (Salpeter) and 3.35, shown as
dashed, solid, and dot-dashed lines, respectively. Right: 
cumulative distributions of the number (thick lines) and of the stellar
mass (thin lines) for the three IMFs with the same line encoding
as in the left panel.} 
\label{imf}
\end{figure}
\begin{figure}[h!t]
\includegraphics[width=8.3cm]{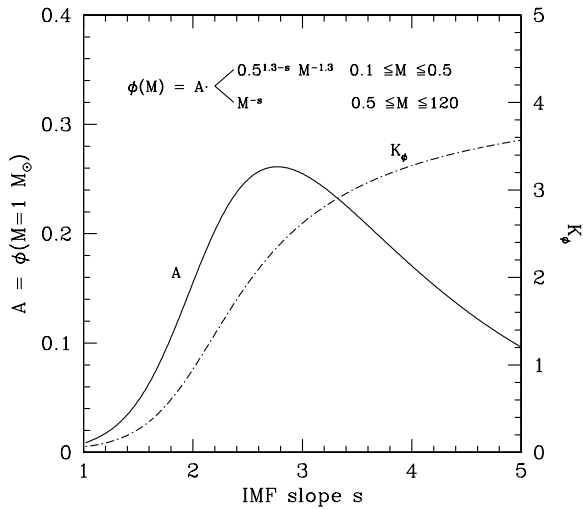} 
\vskip -0.3 truecm
\caption{\footnotesize The scale factor $A$ as a function of the IMF slope. 
All IMFs are normalized to a unitary total mass.
Also shown is the corresponding mass-to-number conversion
factor $K_\phi$, such that the total number of stars is given by
$K_\phi$ times the mass turned into stars (in solar units).}
\label{scalef}
\end{figure}
cumulative  distributions, defined 
as the number of stars with mass less
than $M$, $N(M'<M)=\int_{0.1}^M\phi(M')dM'$, while thin lines show
the fraction of mass in stars less massive than $M$.
In a Salpeter-diet IMF $\sim 0.6\%$ of all stars are more massive than
8 \msun, while for $s=3.35$ and 1.5 these fractions are $0.03\%$
and $9 \%$ respectively. The mass in stars heavier than 8 \msun\ is
20\% for a Salpeter-diet IMF; it drops to 1\% for $s=$
3.35, and is boosted to 77 \% for $s=$ 1.5,  a {\it top-heavy}
IMF. Figure 8.1 wants to convey the message that IMF variations have
a drastic effect on stellar demography, and therefore on several key
properties of stellar populations. Suffice it to say that most of the 
nucleosynthesis comes from $M>10\,\msun$ stars, whereas the light of
an old population (say, $t>10$ Gyr) comes from stars with $M\simeq
1\,\msun$, and therefore is proportional to $\phi(M=\,\msun)$.

Figure~8.2 shows the variation of scale factor $A$
(cf. Chapter 2) as a function of the IMF slope, again for a fixed amount
${\cal M}$ of gas turned into stars.  The scale factor $A$ has a
maximum for $s\simeq 2.75$, pretty close to the Salpeter's slope. Since
by construction $A=\phi(M=1\,\msun)$ and the luminosity of a $\gsim 10$ Gyr old
population is proportional to $\phi(M=1\,\msun)$, an IMF
with the Salpeter's slope has the remarkable property of almost
maximizing the light output of an old population, for fixed mass
turned into stars. A flat IMF ($s=1.35)$ is much less efficient in this respect,
indeed by a factor of $\sim 8$ compared to the Salpeter's slope, as
shown by Figure~8.2. This  figure also shows the
mass-to-number conversion factor $K_\phi$, giving the number of stars $N_{\rm
T}$ formed out of a unit amount of gas turned into stars, that is, $N_{\rm
T}=K_\phi {\cal M}/\msun$. Thus, for a Salpeter-diet IMF $K_\phi\simeq
1.5$, saying that $\sim 150$ stars are formed out of $100\,\msun$ of
gas turned into stars.

An empirically motivated, broken-line IMF such as that shown in
Figure~8.1 is widely adopted in current astrophysical
applications, yet Nature is unlike to make such a cuspy IMF. Perhaps a
more elegant rendition of basically the same empirical data is
represented by a Salpeter+lognormal distribution in which a lognormal
IMF at low masses joins smoothly to a Salpeter IMF at higher masses,
that is:
\begin{equation}
\begin{tabular}{ r l }
\(M\phi(M) $ & \(= A_1\, {\rm exp}\, [-({\rm Log}\, M-{\rm Log}\, 
                    M_{\rm c})^2/2\sigma^2], \quad\quad {\rm for\;\;} M\le 1\,
                    \msun\)
\\
&\( = A_2M^{-x}, \quad\quad\quad\quad\quad\quad\quad\quad\quad\quad
                  \quad\quad\quad\; \;{\rm for\;\;} M> 1\,\msun \)$$
\end{tabular}
\label{eq_cha}
\end{equation}
where $A_1 = 0.159$, $M_{\rm c}=0.079$, $\sigma=0.69$, $A_2=0.0443$
and $x=1.3$. Thus, this {\it Chabrier} IMF is almost identical to the
Salpeter IMF above $1\,\msun$, and smoothly flattens below, being
almost indistinguishable from the Salpeter-diet IMF.

Explorations of variable IMFs can be made by either changing its
slope, or by moving to higher/lower masses the break of the IMF slope
with respect to Equation~(8.1), or allowing the
characteristic mass $M_{\rm c}$ in Equation (8.3) to vary.
Figure~8.3 shows examples of such evolving IMFs.
The two slope IMF with $M_{\rm break}=0.5\,\msun$ and the Chabrier IMF
with $M_{\rm c}=0.079\,\msun$ (lines $a$ and $c$ in Figure~8.3) fit each
other extremely well and both provide a good fit to the local
empirical IMF. By moving the break/characteristic mass to higher
values one can explore the effects of such evolving IMF, for example
mimicking 
\begin{figure}[h!b]
\includegraphics[width=8.7cm]{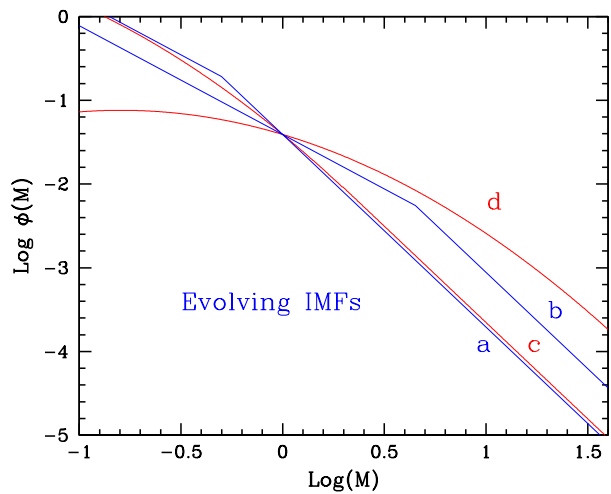} 
\vskip -0.5 truecm
\caption{\footnotesize Examples of evolving IMFs, for a two-slope IMF and a
Chabrier-like IMF. Lines {\it a} and {\it c} represent the local
IMF. The other lines show modified IMFs to explore a hypothetical
evolution with redshift, with the break mass and the characteristic
mass $M_{\rm c}$ having increased to $\sim 4\,\msun$, lines {\it b}
and {\it d} respectively for the two-slope and the Chabrier-like IMF.
All IMFs have been normalized to have the same value for $M=\msun$.}
\label{imfevol}
\end{figure}
a systematic trend with redshift. 
The cases with $M_{\rm
c}\simeq M_{\rm break}\simeq 4\,\msun$ are shown in
Figure~8.3 (lines $b$ and $d$). Having normalized all IMFs to the same value of
$\phi(M=1\,\msun)$, Figure 8.3 allows one to immediately gauge the
relative importance of massive stars compared to solar mass stars,
with the latter ones providing the bulk of the light from old ($\gsim
10$ Gyr) populations.

\vspace{0.65 truecm}
\pn
\textbf{8.2}
\pn
\textbf{The $M/L$  Ratio of Elliptical Galaxies and the IMF 
Slope Below $1\,\msun$}
\vspace{0.5 truecm}
\pn
Figure~8.4 shows as a function of age the $\m*/\lb$ ratio
(where \m* is the stellar mass) for SSPs with solar composition, and different IMFs each with a single
slope $s$ over the whole mass range $0.1 \leq M \leq 100\,\msun$. Very large
mass-to-light ratios are produced by either very flat ($s=1.35$) or
very steep ($s=3.35$) IMFs, whereas the Salpeter's slope gives the
lowest values of the $\m*/\lb$ ratio. This is a result of the
different stellar demography already illustrated in Figures~8.1
and 8.2, such that a steep IMF is \textit{dwarf
dominated}, that is, most of the mass is in low-mass stars, whereas a
flat IMF is \textit{remnant dominated} and most of the mass is in dead
remnants.
\begin{figure}[h!b]
\includegraphics[width=8.7cm, angle=0]{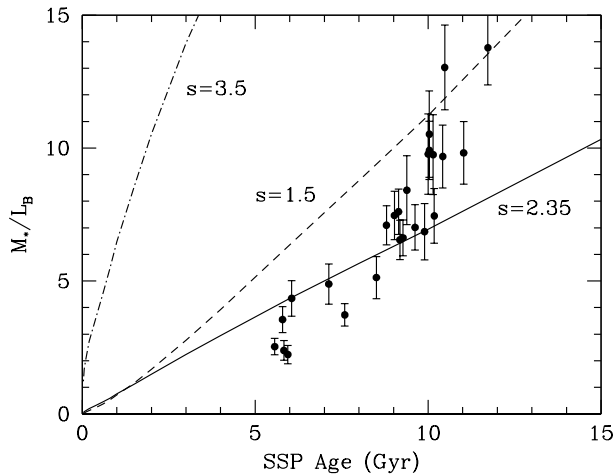} 
\vskip -0.5 truecm
\caption{\footnotesize The stellar mass-to-light ratio of solar metallicity SSPs as
a function of age, for three different single slope IMFs, from 0.1 to
100 $\msun$. Also plotted are values of the dynamical $M/L$ ratio for
a sample of local elliptical galaxies with detailed dynamical
modelling (source: $M/L$ ratios for models: Maraston, C. (1998, {\it
  Mon. Not. R. Astron. Soc.}, 300, 872); for the data: Cappellari,
M. {\it et al.} (2006, {\it
  Mon. Not. R. Astron. Soc.}, 366, 1126), van der Marel, R.P. and van
Dokkum, P. (2007, {\it Astrophys. J.}, 668, 756), van Dokkum, P. and
van der Marel, R.P. (2007,  {\it Astrophys. J.},  655, 30); ages: from
Eq. (1) in Thomas, D. {\it et al.} (2010,  {\it
  Mon. Not. R. Astron. Soc.}, 404, 1775)).}
\label{mtolb98}
\end{figure}
\begin{figure}[h!b]
\includegraphics[width=9cm, angle=0]{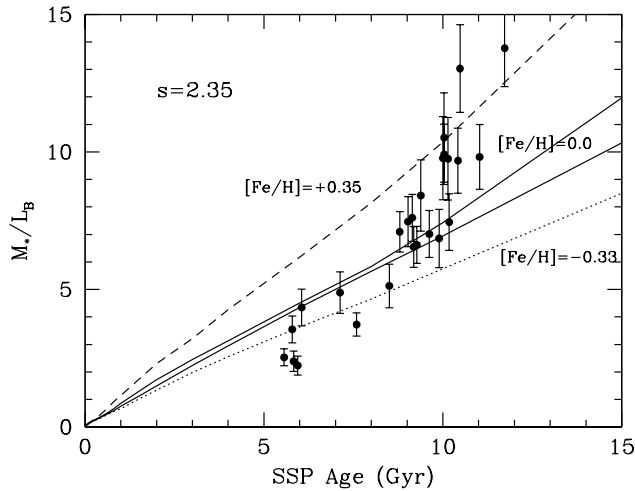} 
\vskip -0.5 truecm
\caption{\footnotesize The $M/L$ ratio of SSPs with a straight Salpeter IMF, for 
subsolar (dotted), supersolar (dashed) and solar metallicity (solid), as
indicated. The two solid lines refer to two releases of the same set
of SSP models (source: model $M/L$ ratios are from Maraston, C. (1998, {\it
  Mon. Not. R. Astron. Soc.}, 300, 872; 2005,  {\it
  Mon. Not. R. Astron. Soc.}, 362, 799); data points are the same as
in Figure 8.4).}
\label{mtolb05}
 \end{figure}

Measurements of the structure (e.g., half light radius) and stellar
velocity dispersion of elliptical galaxies provide estimates of their
{\it dynamical} mass, hence their dynamical mass-to-light ratio can be
compared to the stellar $M/L$ ratio. This is shown in
Figure~8.4 for a sample of local elliptical galaxies with
detailed dynamical modelling,  having adopted a relation between
the luminosity-weighted $\,$ age of their $\,$ stellar populations and velocity
$\,$ dispersion, namely ${\rm Log(Age/Gyr)} =$ $ -0.11 + 0.47\,{\rm
Log}(\sigma_{\rm v})$, consistent with Eq. (6.16). Clearly very steep ($s=3.5$) and very flat
($s=1.5$) slopes of the IMF appear to be excluded by the data, whereas
the intermediate (Salpeter) slope is quite consistent with the data,
apart from the older galaxies which have a higher $M/L$ ratio than the
SSP models. However, besides an increase of age also the average
metallicity is likely to increase with $\sigma_{\rm v}$, with the
galaxies in Figure~8.4 spanning a range from $\sim 1/2$
solar to $\sim 2$ times solar. Thus, the same galaxies are displayed
again in Figure~8.5, together with model $M/L$ ratios for a
straight Salpeter IMF and three different metallicities. The
trend in $M/L$ ratio exhibited by the data appears to be consistent
with the trend resulting from the metallicity trend with
$\sigma_{\rm v}$, and with a straight Salpeter IMF. However, things
may not be as simple as they appear. Dark matter may contribute to the
dynamical $M/L$ ratios, and the IMF may not be straight Salpeter. A
Salpeter-diet IMF such as that shown in Figure~8.1 would give
$\m*/\lb$ ratios systematically lower by $\sim 40\%$ than shown in these
figures, thus opening some room for a dark matter contribution to the
dynamical mass of these galaxies. 
Alternatively, an IMF slightly flatter than Salpeter at high masses,
with its larger contribution by stellar remnants, would reproduce the
high dynamical $M/L$ ratios of the oldest galaxies, without dark
matter contribution. It is quite difficult to circumvent this dark-matter/IMF 
degeneracy on the dynamical $M/L$ ratios of elliptical galaxies. 
\newpage
$\,$\vspace{-1.5 truecm}
\pn
\textbf{8.3}
\pn
\textbf{The Redshift Evolution of the $M/L$ Ratio of Cluster 
Ellipticals and the  IMF Slope Between $\sim 1$ and $\sim 1.4\, M_\odot$}
\vspace{0.5 truecm}
\pn
The slope of the IMF controls the rate of luminosity evolution of a
SSP, as shown by Figure~2.6 for the bolometric light. The flatter the
IMF the more rapid the luminosity declines past an event of star
formation. On the contrary, the steeper the IMF the slower such
decline, as the light from many low-mass stars compensates for the
progressive death of the rarer, more massive and brighter stars.
Having identified and studied passively evolving elliptical galaxies
all the way to $z\sim 2$ and even beyond, one expects that
their $M/L$ ratio must systematically decrease with increasing
redshift, and do so by an amount that depends on the slope of the IMF.
This test is particularly effective if undertaken for cluster
ellipticals, as clusters provide fairly numerous samples of
ellipticals at well defined redshifts. Besides the IMF, the rate of
luminosity ($M/L$) evolution also depends on the age of a SSP, being
much faster at young ages than at late epochs. Thus, the rate of $M/L$
evolution of elliptical galaxies with redshift must depend on both the IMF
slope and the luminosity-weighted age of their stellar populations, or,
equivalently on their formation redshift.

We know that the bulk of stars in local massive ellipticals are very
old, and for an age of $\sim 12$ Gyr the light of such galaxies comes
from a narrow range around the turnoff mass $\mto$ at $\sim
1\,\msun$. Their progenitors at $z\sim 1.5$ must be younger by the
corresponding lookback time, that is, $\sim 9$ Gyr younger, and from the
$\mto-$age relation (Eq. (2.2)) we see that the bulk
of light of such progenitors has to come from stars of mass around
$\mto\simeq 1.4\,\msun$. Thus, the evolution of the $M/L$ ratio of old
stellar populations from redshift zero all the way to redshift $\sim
1.5$ is controlled by the IMF slope in the narrow interval between
$\sim 1$ and $\sim 1.4\,\msun$. The IMF slope below $\sim 1\,\msun$
has no influence on the luminosity evolution, and that above $\sim
1.4\,\msun$ was in control of the luminosity evolution at redshifts
beyond $\sim 1.5$. Therefore, the evolution of the $M/L$ ratio of
elliptical galaxies from $z=0$ to $\sim 1$ allows us to measure the
slope of the IMF just near $M\sim 1\,\msun$.  

Figure~8.6 shows the evolution with redshift of the
$\m*/\lb$ ratio of solar composition SSPs, for various IMF slopes and
different formation redshifts. Also plotted is the average $\m*/\lb$
ratio of cluster ellipticals from the literature, from local clusters
at $z\sim 0$ all the way to clusters at $z\sim 1.3$. A Salpeter slope
($s=2.35$) fits the data for a formation redshift $\zf$ between $\sim
2$ and $\sim 3$, which is in pretty good agreement with both the
formation redshift derived from age-dating local ellipticals in
various ways, and with the observed rapid disappearance of quenched
galaxies beyond $z\sim 2$.  
\begin{figure}[h!t]
\includegraphics[width=8.5cm, angle=0]{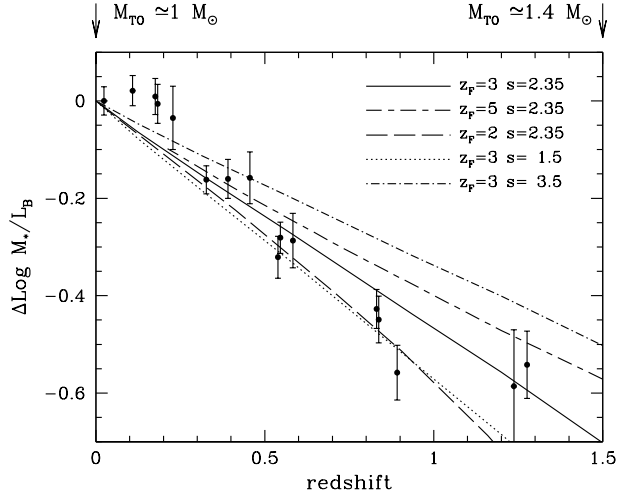} 
\vskip -0.5 truecm
\caption{\footnotesize The differential redshift evolution (with respect to the
value at $z=0$) of the $\m*/\lb$ mass-to-light ratio of solar
composition SSPs, for various choices of the IMF slope between $\sim
1$ and $\sim 1.4\,\msun$, and for various assumed formation redshifts
$\zf$, as indicated. The data points refer to the
$\m*/\lb$ ratio of elliptical galaxies in clusters at various
redshifts, from  $z\sim 0$ up to $z\simeq 1.3$. (Updated from Renzini,
A. (2005) {\it The Initial Mass Function 50 Years later},
(ed. E. Corbelli {\it et al.} , {\it Ap. Sp. Sci. Library}, 327, 221.).}
\label{mtolbold}
\end{figure}
Assuming that the IMF at the formation redshift
of ellipticals was like line {\it b} in Figure~8.3, then
with $s=1.3$ at $M=1\,\msun$ this IMF would require a formation
redshift well beyond 3 in order to fit the data. Line {\it d} instead,
with $s=0.8$ at $M=1\,\msun$ would fail to match the data even
assuming $\zf=\infty$, as shown in Figure~8.6.  One can
conclude that the evolution of the $M/L$ ratio of cluster elliptical
galaxies up to redshift $\sim 1.3$ does not favor any
significant departure from the Salpeter value $s=2.35$ of the slope of
the IMF in the vicinity of $M\sim 1\,\msun$, all the way to a
formation redshift beyond $\sim 2$.

\vspace{0.35 truecm}
\pn
\textbf{8.4}
\pn
\textbf{The Metal Content of Galaxy Clusters and the IMF Slope
 Between $\sim 1$ and $\sim 40\,\msun$, and Above}
\vspace{0.5 truecm}
\pn
In its youth, a stellar
population generates lots of UV photons, core collapse supernovae and
metals that go to enrich the ISM. In its old age, say $\sim 12$ Gyr later,
the same stellar population radiates optical-near-IR light from its
$\sim 1\,\msun$ stars, while all more massive stars are dead remnants.
The amount of metals ($M_{\rm X}$) that are produced by such
populations is proportional to the number of massive stars $M\gsim
8\,\msun$ that have undergone a core collapse supernova explosion,
whereas the luminosity (e.g., $\lb$) at $t\simeq 12$ Gyr is
proportional to the number of stars with $M\sim 1\,\msun$. It follows
that the metal-mass-to-light ratio $M_{\rm X}/\lb$ is a measure of
the number ratio of massive to $\sim\msun$ stars, that is, of the IMF
slope between $\sim 1$ and $\sim 40\,\msun$. 
Clusters of galaxies offer an excellent opportunity to measure both
the light of their dominant stellar populations, and the amount of
metals that such populations have produced in their early days. 
Indeed, most of the light of clusters of galaxies comes from $\sim 12$
Gyr old, massive ellipticals, and the abundance of metals can be
measured both in their stellar populations and in the intracluster
medium (ICM).  Iron is the element whose abundance can be most
reliably measured both in cluster galaxies and in the ICM, but its
production is likely to be dominated by Type Ia supernovae whose
progenitors are binary stars. As extensively discussed in Chapter 7,
a large fraction of the total iron production comes from Type Ia
supernovae, and the contribution from CC supernovae is uncertain;
therefore the $\mfe/\lb$ ratio of clusters is less useful to set constraints on the
IMF slope between $\sim 1$ and $M\gsim 10\,\msun$. For this reason, we
focus on oxygen and silicon, whose production is indeed dominated by
core collapse supernovae.

Following the notations in Chapter 2, the IMF can be
written as:
\begin{equation}
\phi(M) = a(t,Z)\lb M^{-\rm s},
\end{equation}
where $a(t,Z)$ is the relatively slow function of SSP age and
metallicity shown in Figure~2.10, multiplied by the bolometric correction 
shown in Figure~3.1.  Thus, the metal-mass-to-light
ratio for the generic element ``X '' can be readily calculated from:
 \begin{equation}
{M_{\rm X} \over\lb}={1 \over \lb} \, \int_8^{120}m_{\rm X}(M)\phi(M)dM = a(t,Z)\int_8^{120}
m_{\rm X}(M)M^{-\rm s}dM,
\label{mxlb}
\end{equation}
\pn
where $m_{\rm X}(M)$ is the mass of the element X which is produced
and ejected by a star of mass $M$.  From stellar population models one
has $a(12\,{\rm Gyr},Z)=2.22$ and 3.12, respectively, for $Z=\zsun$ and
$2\zsun$ and we adopt $a(12\,{\rm Gyr})=2.5$ in Eq.~(8.5). Using
the oxygen and silicon yields $m_{\rm O}(M)$ and $m_{\rm Si}(M)$ from
theoretical nucleosynthesis (cf. Figure~7.5), Equation~(8.5)
then gives the $M_{\rm O} /\lb$ and $M_{\rm Si} /\lb$ metal mass-to-light
ratios that are reported in Figure~8.7 as a function of the
IMF slope between $\sim 1$ and $\sim 40\,\msun$. As expected, the
$M_{\rm O} /\lb$ and $M_{\rm Si} /\lb$ are extremely sensitive to the
IMF slope. The values observed in local clusters of galaxies, from
X-ray observations of the ICM and assuming stars are near solar on
average, are $\sim 0.1$ and $\sim 0.008\,\msun/\lsun$, respectively
for oxygen and silicon, as documented in Chapter 10. These empirical
values are also displayed in Figure~8.7. A comparison
with the calculated values shows that with the Salpeter IMF slope
($s=2.35$) the standard explosive nucleosynthesis from core collapse
supernovae produces just the right amount of oxygen and silicon to
match the observed $M_{\rm O} /\lb$ and $M_{\rm Si} /\lb$ ratios in
clusters of galaxies, having assumed that most of the $B-$band light
of clusters comes from $\sim 12$ Gyr old populations. Actually,
silicon may be even somewhat overproduced if one allows a $\sim 40\%$
contribution from Type Ia supernovae (cf. Figure~7.17).

Figure~8.7 also shows that with $s=1.35$ such a top heavy
IMF would overproduce oxygen and silicon by more than a factor $\sim
20$. Such a huge variation with $\Delta s=1$ is actually expected,
given that for a near Salpeter slope the typical mass of metal
producing stars is $\sim 25\,\msun$. By the same token, the IMF
labelled {\it b} in Figure~8.3 would overproduce metals by a
factor $\sim 4$ with respect to a Salpeter-slope IMF (lines {\it a}
and {\it c} ), whereas the IMF labelled {\it d} in
Figure~8.3 would do so by a factor $\sim 20$. Thus, under
the assumptions that 
\begin{figure}[h!t]
\includegraphics[width=10cm, angle=0]{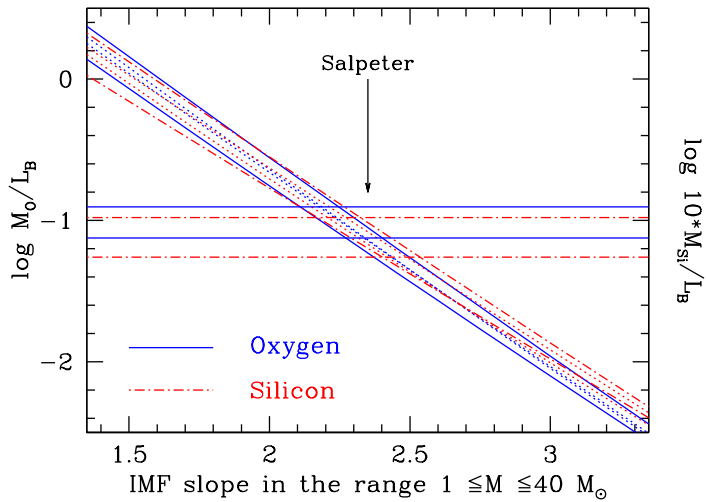} 
\caption{\footnotesize The oxygen and silicon mass-to-light ratios as a function of
the IMF slope for a $\sim 12$ Gyr old, near solar metallicity SSP with
oxygen and silicon yields from standard nucleosynthesis
calculations. Different lines refer to the different theoretical yields 
shown in Figure~7.5a,b. The horizontal lines show the uncertainty 
range of the observed values of these ratios in clusters of galaxies, with
central values as reported in Chapter 10, and allowing for a $\sim \pm
25\%$ uncertainty.  }
\label{spineto6}
\end{figure}
the bulk of the light of local galaxy clusters comes
from $\sim 12$ Gyr old stars, that current stellar theoretical
nucleosynthesis is
basically correct, and that no large systematic errors affect the
reported empirical values of the $M_{\rm O} /\lb$ and $M_{\rm Si}
/\lb$ ratios, then one can exclude a significant evolution of the IMF
with cosmic time, such as for example, one in which the IMF at $z\sim 3$
would be represented by line {\it b} or {\it d} in
Figure~8.3, and by line {\it a} or {\it c} at $z=0$.

\medskip\pn
\begin{center}
------------OOO------------
\end{center}
\medskip\pn 
A variable IMF is often invoked as an \textit{ad hoc} fix to specific
discrepancies that may emerge here or there, which however may have
other origins. For example, an evolving IMF with redshift has been
sometimes invoked to ease a perceived discrepancy between the cosmic
evolution of the stellar mass density, and the integral over 
cosmic time of the star formation rate. In other contexts it has been
proposed that the IMF may be different in starbursts as opposed to
more steady star formation, or in disks vs. spheroids. Sometimes one
appeals to a top-heavy IMF in one context, and then to a bottom-heavy
one in another, as if it was possible to have as many IMFs as problems
to solve. Honestly, we do not know whether there is one and only one
IMF. However, if one subscribes to a different IMF to solve a single
problem, then at the
same time one should make sure the new IMF does not destroy
agreements elsewhere, or conflicts with other
astrophysical constraints.
While it is perfectly legitimate to contemplate IMF
variations from one situation to another, it should be mandatory to
explore all consequences of postulated variations, well beyond the
specific case one is attempting to fix. This kind of sanitary check is
most frequently neglected in the literature appealing to IMF
variations. A few examples of such checks have been presented in
this chapter.

\newpage
$\,$

\par\noindent
{\bf Further readings}

\vspace{5 truemm}\pn
This chapter expands and updates the paper:  Renzini,
A. (2005) {\it The Initial Mass Function 50 Years later},
ed. E. Corbelli {\it et al.} , Ap. Sp. Sci. Library, 327, 221.

\bigskip\pn {\bf Most Popular Initial Mass Functions} 

\medskip\pn
Chabrier, G. (2003) {\it Publ. Astron. Soc. Pac.}, {\bf 115}, 763.\\
Kroupa, P. (2002) {\it Science}, {\bf 295}, 82.\\
Salpeter, E.E. (1955) {\it Astrophys. J.},  {\bf 121}, 161.\\
Scalo, J.M. (1986) {\it Fund. Cosm. Phys.}, {\bf 11}, 1.

\bigskip 
\pn {\bf Recent review} 
\medskip\pn
Bastian, N. et al. (2010) {\it Annu. Rev. Astron. Astrophys.}, {\bf 48}, 339.

\bigskip 
\pn {\bf The IMF at High Redshift} 

\medskip\pn
Dav\'e, R. (2008) {\it Mon. Not. R. Astron. Soc.}, {\bf 385}, 147.\\
Tacconi, L.J. (2008) {\it Astrophys. J.}, {\bf 680}, 246.\\
van Dokkum, P. G. (2008) {\it Astrophys. J.}, {\bf 674}, 29.

\newpage
$\,$\pn
{\bf Table of Contents of the whole book}\\
\\
{\bf 1 Firm and Less Firm Outcomes of Stellar Evolution Theory} \hfill 1 \\
1.1 A Brief Journey through Stellar Evolution \hfill 1 \\
1.1.1 A $9\;\msun$  Star \hfill 1 \\
1.1.2 The Evolution of Stars with Solar Composition \hfill 8 \\
1.1.3 Dependence on Initial Chemical Composition \hfill 12 \\
1.1.4 The Asymptotic Giant Branch Phase \hfill 15 \\
1.2 Strengths and Weaknesses of Stellar EvolutionaryModels \hfill 18 \\
1.2.1 Microphysics \hfill 19 \\
1.2.2 Macrophysics \hfill 20 \\
1.3 The Initial Mass-Final Mass Relation \hfill 31 \\
\\
{\bf 2 The Fundamentals of Evolutionary Population Synthesis} \hfill 35 \\
2.1 The Stellar Evolution Clock \hfill 35 \\
2.2 The Evolutionary Flux \hfill 38 \\
2.3 The Fuel Consumption Theorem \hfill 39 \\
2.4 Fuel Consumptions \hfill 42 \\
2.5 Population Synthesis Using Isochrones \hfill 46 \\
2.6 The Luminosity Evolution of Stellar Populations \hfill 47 \\
2.7 The Specific Evolutionary Flux \hfill 49 \\
2.8 The IMF Scale Factor \hfill 51 \\
2.9 Total and Specific Rates of Mass Return \hfill 52 \\
2.10 Mass and Mass-to-Light Ratio \hfill 56 \\
2.11 IMF-Dependent and IMF-Independent Quantities \hfill 57 \\
2.12 The Age-Metallicity Degeneracy \hfill 58 \\
\\
{\bf 3 Resolving Stellar Populations} \hfill 61 \\
3.1 The Stellar Populations of Pixels and Frames \hfill 61 \\
3.1.1 The Stellar 1 Population of a Frame \hfill 61\\
3.1.2 The Stellar Population of a Pixel \hfill 64\\
3.2 Simulated Observations and Their Reduction\hfill  68\\
\\
{\bf 4 Age Dating Resolved Stellar Populations} \hfill 77\\
4.1 Globular Cluster Ages \hfill 77\\
4.1.1 Absolute and Relative Globular Cluster Ages \hfill 78\\
4.1.2 Globular Clusters with Multiple Populations \hfill 80\\
4.2 The Age of the Galactic Bulge \hfill 83\\
4.3 Globular Clusters in the Magellanic Clouds \hfill 86\\
4.4 Stellar Ages of the M31 Spheroid \hfill 88\\
4.4.1 The Bulge of M31 \hfill 88\\
4.4.2 The M31 Halo and Giant Stream \hfill 90\\
4.5 The Star Formation Histories of Resolved Galaxies \hfill 92\\
4.5.1 The Mass-Specific Production \hfill 93\\
4.5.2 Decoding the CMD \hfill 98\\
4.5.3 The Specific Production Method \hfill 102\\
4.5.4 The Synthetic CMD Method \hfill 104\\
4.5.5 An Example: the Stellar Population in the Halo of the Centaurus A

 Galaxy \hfill 106\\
\\
{\bf 5 The Evolutionary Synthesis of Stellar Populations} \hfill 113\\
5.1 Simple Stellar Populations \hfill 113\\
5.2 Spectral Libraries \hfill 115\\
5.2.1 Empirical Spectral Libraries \hfill 115\\
5.2.2 Model Atmosphere Libraries \hfill 116\\
5.3 Composite Stellar Populations \hfill 116\\
5.4 Evolving Spectra \hfill 118\\
5.4.1 The Spectral Evolution of a SSP \hfill 118\\
5.4.2 The Spectral Evolution of Composite Stellar Populations \hfill 121\\
5.4.3 There are Also Binaries \hfill 128\\
\\
{\bf 6 Stellar Population Diagnostics of Galaxies} \hfill 133\\
6.1 Measuring Star Formation Rates \hfill 133\\
6.1.1 The SFR from the Ultraviolet Continuum \hfill 134\\
6.1.2 The SFR from the Far-Infrared Luminosity \hfill 136\\
6.1.3 The SFR from Optical Emission Lines \hfill 137\\
6.1.4 The SFR from the Soft X-ray Luminosity \hfill 138\\
6.1.5 The SFR from the Radio Luminosity \hfill 139\\
6.2 Measuring the Stellar Mass of Galaxies \hfill 140\\
6.3 Age and Metallicity Diagnostics \hfill 143\\
6.3.1 Star-Forming Galaxies \hfill 143\\
6.3.2 Quenched Galaxies \hfill 145\\
6.4 Star-Forming and Quenched Galaxies through Cosmic Times \hfill 153 \\
6.4.1 The Main Sequence of Star-Forming Galaxies \hfill 155 \\
6.4.2 The Mass and Environment of Quenched Galaxies \hfill 163 \\
6.4.3 Mass Functions \hfill 164 \\
\\
{\bf 7 Supernovae} \hfill 171 \\
7.1 Observed SN Rates \hfill 173 \\
7.2 Core Collapse SNe \hfill 175 \\
7.2.1 Theoretical Rates \hfill 176 \\
7.2.2 Nucleosynthetic Yields \hfill 179 \\
7.3 Thermonuclear Supernovae \hfill 184 \\
7.3.1 Evolutionary Scenarios for SNIa Progenitors \hfill 185 \\
7.3.2 The Distribution of Delay Times \hfill 187 \\
7.3.3 The SD Channel \hfill 188 \\
7.3.4 The DD Channel \hfill 191 \\
7.3.5 Constraining the DTD and the SNIa Productivity \hfill 197 \\
7.3.6 SNIa Yields \hfill 201 \\
\\
\newpage
$\,$\pn
{\bf 8 The IMF From low to High Redshift} \hfill 207 \\
8.1 How the IMF Affects Stellar Demography \hfill 208 \\
8.2 The $M/L$ Ratio of Elliptical Galaxies and the IMF Slope Below 

1 $\msun$ \hfill  211 \\
8.3 The Redshift Evolution of the $M/L$ Ratio of Cluster Ellipticals 

and the  IMF Slope Between $\sim 1$ and $\sim 1.4\;\msun$ \hfill 213 \\
8.4 The Metal Content of Galaxy Clusters and the IMF Slope Between 

$\sim 1$ and $\sim 40\;\msun$, and Above \hfill 214 \\
\\
{\bf 9 Evolutionary Links Across Cosmic Time: an Empirical 

History of Galaxies} \hfill 219 \\
9.1 The Growth and Overgrowth of Galaxies \hfill 221 \\
9.2 A Phenomenological Model of Galaxy Evolution \hfill 224 \\
9.2.1 How Mass Quenching Operates \hfill 225 \\
9.2.2 How Environmental Quenching Operates \hfill 227 \\
9.2.3 The Evolving Demography of Galaxies \hfill 229 \\
9.2.4 Caveats \hfill 232 \\
9.2.5 The Physics of Quenching \hfill 234 \\
\\
{\bf 10 The Chemical Evolution of Galaxies, Clusters, and the 

Whole  Universe} \hfill 237 \\
10.1 Clusters of Galaxies \hfill 237 \\
10.1.1 Iron in the Intracluster Medium and the Iron Mass-to-Light Ratio \hfill 238 \\
10.1.2 The Iron Share Between ICM and Cluster Galaxies \hfill 244 \\
10.1.3 Elemental Ratios \hfill 245\\
10.1.4 Metal Production: the Parent 1 Stellar Populations \hfill 247\\
10.1.5 Iron from SNIa \hfill 248\\
10.1.6 Iron and Metals from Core Collapse SNe \hfill 249\\
10.2 Metals from Galaxies to the ICM: Ejection versus Extraction \hfill 250\\
10.3 Clusters versus Field and the Overall Metallicity of the Universe \hfill 252\\
10.4 Clusters versus the Chemical Evolution of the Milky Way \hfill
254\\
\end{document}